# Secrecy Performance of Terahertz Wireless Links in Rain and Snow

Rong Wang[1#], Yu Mei[1#], Xiangzhu Meng[2], Jianjun Ma[1*]

[1]School of Information and Electronics, Beijing Institute of Technology, Beijing 100081, China
[2]School of Automation, Beijing Institute of Technology, Beijing 100081, China

[#] Rong Wang and Yu Mei contributed equally to this work.
[*]Correspondence: jianjun_ma@bit.edu.cn

**Abstract**

Wireless communication technique operating at terahertz (THz) frequencies is regarded as the most potential candidate for future wireless networks due to its wider frequency bandwidth and higher directionality when compared with that employing radio frequency (RF) and millimeter waves (mmWaves). It has been proved that the high directionality owned by THz wireless links could help to reduce the possibility of eavesdropping attacks at physical layer. However, for outdoor applications in adverse weathers (such as water fog, dust fog, rain and snow), scattering and absorption effects suffered by a THz link due to weather particles and gaseous molecules could degrade its secrecy performance seriously. In this work, we present theoretical investigations on physical layer security of a point-to-point THz link in rain and snow with a potential eavesdropper locating outside of the legitimate link path. Signal degradation due to rain/snow, gaseous attenuation and beam divergence are included in a theoretical model to estimate the link performance. Secrecy capacity of the link with carriers at 140, 220 and 340 GHz is calculated and compared. We find that the rain/snow intensity, carrier frequency and receiver sensitivity could affect the secrecy performance and their influence on insecure region and maximum safe data transmission rate is discussed and summarized.

## 1. Introduction

Unlike wire communications, the broadcasting nature of wireless links could make it to be vulnerable to eavesdropping attacks at physical layer. One alternative solution for that is updating carrier frequencies to terahertz (THz) range with higher beam directionality and narrower beam width [1]. This makes it to be much harder and even impossible to intercept legitimate signals. Terahertz communication technique could offer such a solution and achieve higher physical layer security when compared with that employing radio frequency (RF) and millimeter waves (mmWaves). This has been proved by a line-of-sight (LOS) link with carriers at 100, 200 and 300 GHz when an eavesdropper appears between two legitimate peers and tries to eavesdrop the message in clear weather [2]. However, at the case of adverse weather conditions, due to the serious scattering effect by weather particles and gaseous molecules [3-6], it may be necessary to investigate the possibility of signal eavesdropping by a non-line-of-sight (NLOS) path scattered away the legitimate link path.

There have been lots of efforts concentrated on wireless link secrecy performance in indoor/outdoor scenarios with carriers at microwave [7], millimeter wave [8], terahertz [2], optical [9] and even infrared frequencies [10]. MIMO configuration [11], fading-resistance technique [12] and backscattering parameter [2] were proved to be valuable methods to reduce the possibility of eavesdropping attacks. However, there is still no publication relating to secrecy performance of outdoor THz links, which is definitely required for the future applications of THz communication technique. In this work, considering our previous researches on THz link performance in adverse weathers, we propose a further study on the link secrecy performance in rain and snow, and find out the influencing mechanisms of scattering and absorption effects on it.

## 2. Attenuation by Falling Rain and Snow

The description of link performance in rain and snow requires detailed physical properties including particle shape, dielectric constant, and raindrop/snowdrop size distributions, which have been demonstrated in several publications [4, 6, 13, 14]. In this manuscript, we would not spend more words on it and would get to the theoretical models directly.

Mie scattering theory could be used to calculate the signal loss due to absorption and scattering by rain/snow particles when the ordinary size of particles in the range from mm to cm [15], which is usually comparable to or larger than the THz wavelength. For very small particles, solutions given by the Mie theory also approach the Rayleigh approximation. This method systematically describes the scattering mechanism of THz waves by particles of various sizes in the atmosphere. Under the assumption of single scattering and scatter independence, the attenuation suffered by a THz wave traveling in rain/snow can be obtained [16] by

$$\alpha = 4.343 \cdot 10^3 \int_0^\infty \sigma_{ext}(m, r, \lambda) N(r) dr, \quad (1)$$

where $m$ is the refractive index of the particle and $N(r)$ is the rain/snow size distribution we mentioned before, with $r$ as the radius of melted snow particles. $\sigma_{ext}$ is extinction cross section, which is a function of $m$, $r$ and carrier wavelength $\lambda$. The relationship exists between the absorption cross-section $\sigma_{abs}(m, r, \lambda)$ and the scattering cross-section $\sigma_{sca}(m, r, \lambda)$ as $\sigma_{ext}(m, r, \lambda) = \sigma_{abs}(m, r, \lambda) + \sigma_{sca}(m, r, \lambda)$, because the signal attenuation by rain/snow is mainly due to absorption and scattering effects.

The transmission of THz waves in rain/snow also suffers attenuation due to gaseous molecules (such as oxygen and water vapor) in atmosphere. Because of the small volume of gaseous molecules, the scattering effect could be neglected over a short distance, but the absorption effect is obvious and should always be considered. Here, we would employ a theoretical model provided by the ITU (International Telecommunication Union) Recommendation Sector (ITU-R) [17] based on the physical model MPM93 [18]. This method is proved to be valid at frequencies below 500 GHz [19] and has been successfully used in our previous publication [6].

## 2.1 Signal Degradation in Falling Rain

The attenuation suffered by THz waves propagating in falling rain is considered in the near-surface atmosphere, which is difficult to analyze due to the physical nature of the rain particles, as there are a lot of weather dependent variations in type, shape, dielectric constants and size distribution [20]. In addition to the Mie scattering theory, a specific attenuation model proposed by ITU-R could also give relatively good average results. But it does not include the effect of raindrop/snowdrop size distribution, which could produce obvious differences even under the same rainfall rate. So we would employ the Mie scattering theory to estimate the link degradation under different raindrop size distribution models, which has been used in [21] to predict THz spectral attenuation and the result is in good agreement with measured data.

The spectral attenuation due to rain is shown in Fig. 1 with several commonly used raindrop size distribution models (Marshall-Palmer [22], JOSS [23] and Weibull [24]) considered. We set temperature and pressure at 25ºC and 1013 hPa, respectively. Relative humidity (RH) in the air is set at 97% as in Table 1. The total attenuation combining Mie scattering theory and gaseous attenuation is calculated. We can see that the total attenuation in rain increases at higher frequencies due to large gaseous absorption. When no gaseous attenuation included, this curve should keep at the same level for frequencies above 100 GHz as measured in [4]. We could also observe that the signal loss due to scattering is much smaller than that due to absoprtion because of much water contined in rain. The scattering loss does not change for freqencies above 100 GHz, which makes the difference between losses due to absorption and scattering increases significantly with the incresing of carrier frequencies. When we compare the signal degradation under three different raindrop size distributions, same evolution in total attenuation, absorption and scattering could be observed, which indicates that the THz link under such rain conditions would own identical or almost identical link performance.

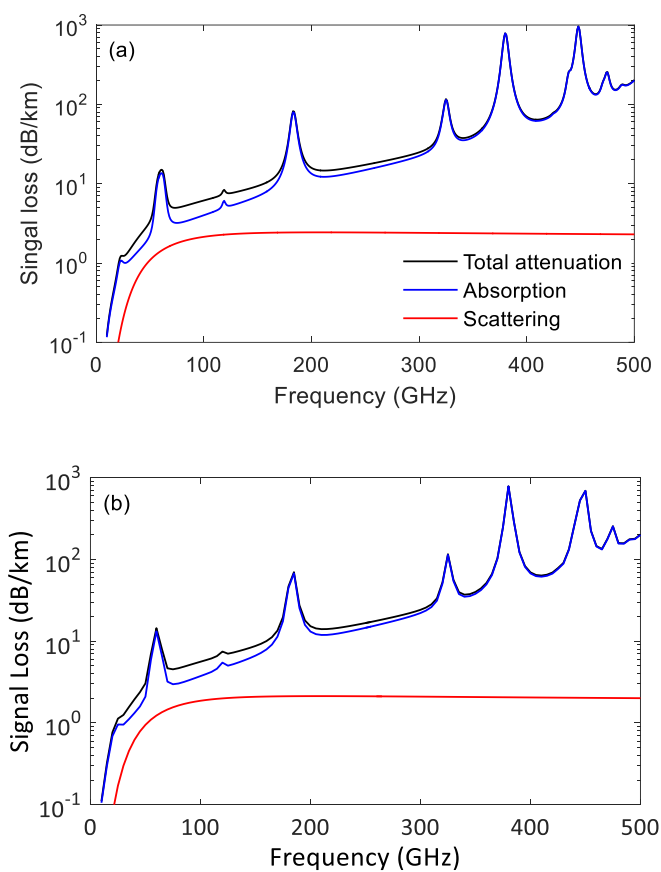

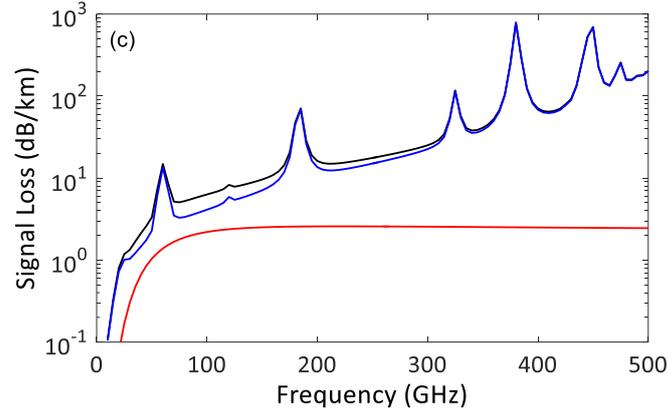

**Figure 1** Attenuation by rain based on Eq. (1) under a rainfall rate of 10 mm/hr (heavy rain [25]) when (a) Marshall-Palmer, (b) Joss and (c) Weibull distributions are employed. ($T$ = 25ºC, $P$ =1013 hPa and RH = 97%)

Table 1 List of parameters employed in all the calculations

| Parameter | Value |
| --- | --- |
| Location of Alice | (0m, 0m) |
| Location of Bob | (1km, 0m) |
| Location of Eve | (100m, 10m) |
| Temperature ($T$) | 25ºC in rain<br>-1ºC in dry snow<br>0ºC in wet snow |
| Pressure ($P$) | 1013 hPa |
| Relative humidity (RH) | 97% |
| Volumetric water content ($m_v$) | 25% for wet snow |
| Receiving area | 1cm$^2$ for Bob<br>1cm$^2$ for Eve |
| Receiver sensitivity by SNR | 0dB for Bob<br>0dB for Eve |

## 2.2 Signal Degradation in Falling Snow

The total signal attenuation by dry and wet snow is calculated by employing Mie scattering theory with Gunn-Marshall (G-M) and Sekhon-Srivastava (S-S) snowdrop size distributions considered. Detailed distribution parameter definition should be found in [6]. Dry snow can be treated as a mixture of ice and air, and its dielectric properties could be obtained by an empirical formula in [26]. Atmospheric pressure is set at 1013 hPa, relative humidity (RH) in the air is 97 % and snowfall rate (equivalent rainfall rate) is 10 mm/hr (Here, over the whole manuscript, we use equivalent rainfall rate to represent the snowflake fall rate as what we demonstrated previously in [10]). For dry snow in

Fig. 2(a) and Fig. 2(c) when temperature $T$ = -1°C, the attenuation by scattering (red curve) on THz waves is close to the total attenuation (black curve, consisting of absorption and scattering) at lower frequencies because there is almost no absorption by dry snow particles [6] with a dielectric constant of 3.1884 over the frequency range.

The total attenuation increases significantly at higher carrier frequencies due to the large gaseous absorption it suffered. The scattering loss also increases, but it becomes smaller than that due to absorption and the difference between both becomes wider.

In the calculation for wet snow as shown in Fig. 2(b) and (d), the parameter settings are identical to that in Fig. 2(a) and (c) except a 0°C temperature. A commonly used Debye-like model is proposed to obtain the complex dielectric constant of wet snow (which is a mixture of ice, air and free water) and was proved in good agreement with measurements in other publications [27]. We can see that the large amount of water contained in wet snow makes the absorption loss always higher than the scattering loss, which would reduce the contribution of scattering to the total attenuation. Like that in rain, the scattering loss in wet snow keeps almost identical for frequencies from 100 GHz to 500 GHz. This means the scattered power by snow would not be frequency-dependent in this frequency range.

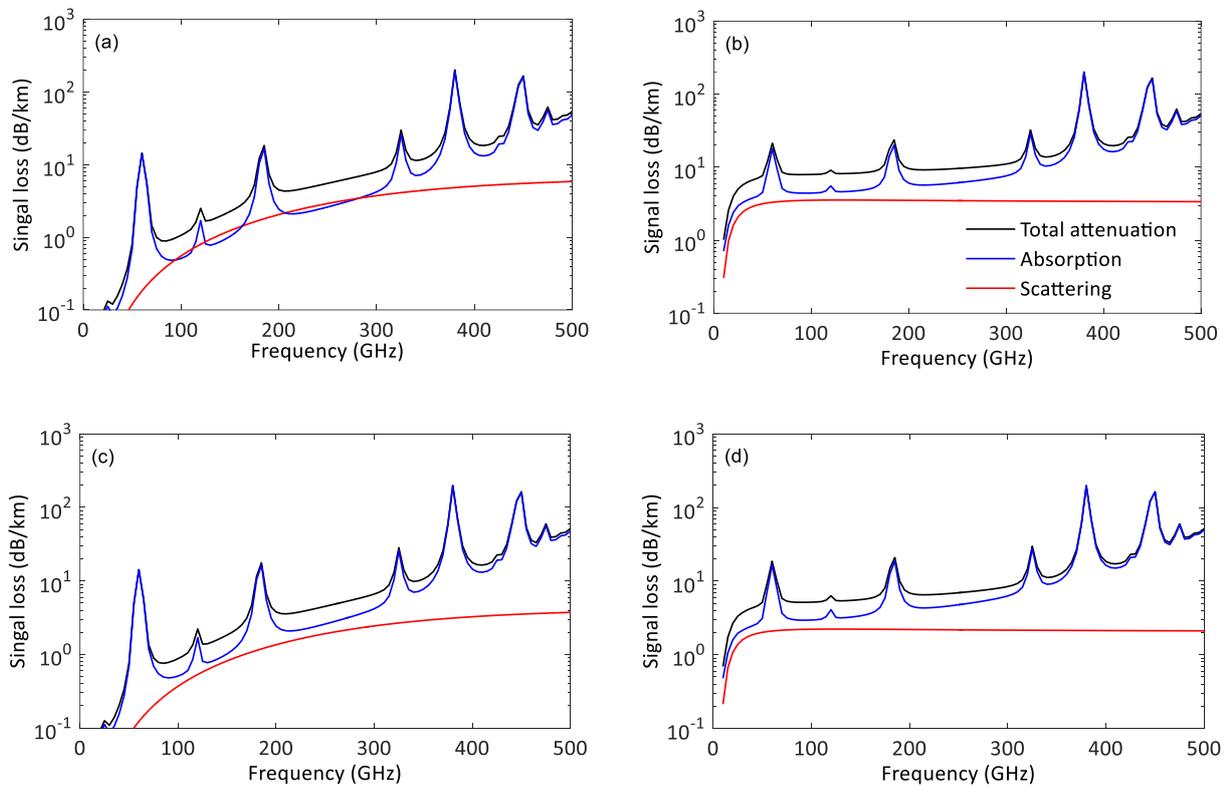

**Figure 2** Total attenuation (black), absorption (blue) and scattering (red) based on Eq. (1) by dry

snow at -1°C under (a) G-M and (c) S-S distributions, and by wet snow (wetness 25%) at 0°C under (b) G-M and (d) S-S distributions with a snowfall rate (equivalent rainfall rate) of 10 mm/hr. (*P*=1013hPa, RH=97%)

## 3. Link Secrecy Performance Analysis

In this section, we assume there is a point-to-point outdoor THz link with the configuration as shown in Fig.3 (a). A transmitter (Alice) sends information to a legitimate receiver (Bob) by a LOS link which suffers absorption and scattering losses due to rain/snow. Meanwhile, an eavesdropper (Eve) exists outside but near the legitimate link path and aims to capture the information through a NLOS channel scattered away from the legitimate (LOS) path. The link distance *d* is set at *d*=1km in all the calculations and the positions of Alice and Bob are always fixed as shown in Table 1, while Eve can change its position and adjust its pointing direction to obtain optimal received signal.

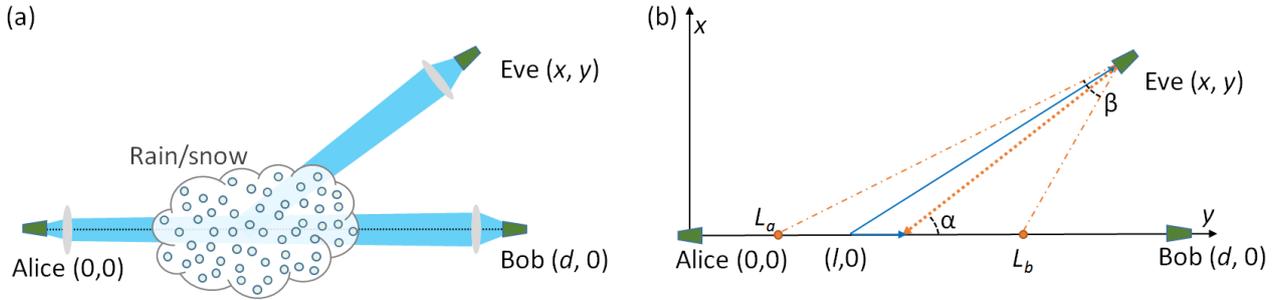

**Figure 3** (a) Geographic of the point-to-point THz link with a security attacker (Eve) in rain or snow; (b) Coordinate system of the legitimate (LOS) and eavesdropping (NLOS) links.

In the presence of rain/snow, the LOS link suffers atmospheric attenuation $G_A$ and divergence attenuation $G_D = 4A/(\pi d^2 \alpha_A^2)$ with $\alpha_A$ as the full divergence angle of Alice and $A$ as the effective receiving area of Bob. The atmospheric attenuation including attenuation only by rain/snow (with coefficient as $\alpha_t$) and gaseous attenuation by air (with coefficient as $\alpha_g$) could be obtained from the Section 2. So the atmospheric attenuation coefficient is $\alpha_{atm} = \alpha_t + \alpha_g$ and the atmospheric attenuation can be obtained by $G_A = \exp(-\alpha_{atm}d)$.

Combining both atmospheric attenuation and divergence attenuation, we can get the total LOS link gain as

$$G_{LOS} = G_A G_D = \frac{4A\exp(-\alpha_{atm}d)}{\pi d^2 \alpha_A^2} \quad . \tag{2}$$

To obtain the link gain of the eavesdropping (NLOS) link achieved by scattering, a widely used model, called single-scattering model [28, 29], is employed. In the link geometry shown in Fig. 1(b), we set Alice at the origin of the coordinate (0, 0), Bob at (d, 0) and Eve at (x, y) with x and y as variables. The signal is transmitted along the positive direction of the x-axis between Alice and Bob. The NLOS link gain $G_{NLOS}$ can be obtained as in [30] by

$$G_{NLOS} = \int_{L_a}^{L_b} \Omega(l) p(\mu) \alpha_{atm} \exp\left\{-\alpha_{atm}[l + \sqrt{(x-l)^2 + y^2}]\right\} dl, \quad (3)$$

with integral variable $l$ as the signal transmission distance before scattering occurs. The lower bound $L_a$ and upper bound $L_b$ can be expressed as

$$L_a = \min\left\{\max\left\{x - \frac{y}{\tan(\alpha - \beta/2)}, 0\right\}, d\right\} \quad (4)$$

and

$$L_b = \min\left\{\max\left\{x - \frac{y}{\tan(\alpha + \beta/2)}, 0\right\}, d\right\}, \quad (5)$$

which divides the scattering region. Here, $\alpha$ is the angle between the pointing direction of Eve and the positive x axis. $\beta$ is the field-of-view (FOV) full angle of Eve. We assume that Eve tries to optimize the pointing direction to maximize the link gain of the NLOS channel while its position and $\beta$ are fixed. So the optimal scattering angle, corresponding to maximum NLOS link gain, can be expressed as $\alpha^* = \arctan(y/x) + \beta/2$ for $\alpha < \pi/2 - \beta/2$ [31]. $\Omega(l)$ denotes the solid angle from receiving area to the scattering point as

$$\Omega(l) = \frac{A}{\left[(x-l)^2 + y^2\right]^{3/2}} \frac{(x-l) + y\tan\alpha}{\sqrt{1 + \tan^2\alpha}}. \quad (6)$$

The phase function $p(\mu)$ can be expressed as

$$p(\mu) = \frac{1-g^2}{4\pi}\left[\frac{1}{(1+g^2 - 2g\mu)^{3/2}} + f\frac{3\mu^2 - 1}{2(1+g^2)^{3/2}}\right], \quad (7)$$

where generalized Henyey-Greenstein function is adopted, and it is defined as scattering phase function indicating the probability distribution of scattering angle. $\mu = (x-l)/\sqrt{(x-l)^2 + y^2}$ represents the cosine of scattering angle in $(l, 0)$ with g as an asymmetry factor related to wavelength, scattering particle radius and refractive index [32].

Secrecy capacity is defined as the maximum data rate from Alice to Bob when perfect secrecy performance is maintained [33] and can be expressed as

$$C_s = [I(X;Y) - I(X;Z)]^+, \quad (8)$$

with $[x]^+ = \max\{0, x\}$ indicating that the secrecy capacity can never be less than 0. Parameters *X*, *Y* and *Z* represents the signals of the Alice, Bob and Eve, respectively. *I(X;Y)* and *I(X;Z)* denote the mutual information of LOS and NLOS links [34], respectively, and can be expressed as

$$I(X;Y) = q(\lambda_L + \lambda_b)\log(\lambda_L + \lambda_b) + \lambda_b \log(\lambda_b) - (q\lambda_L + \lambda_b)\log(q\lambda_L + \lambda_b) \tag{9}$$

and

$$I(X;Z) = q(\lambda_N + \lambda_b)\log(\lambda_N + \lambda_b) + \lambda_b \log(\lambda_b) - (q\lambda_N + \lambda_b)\log(q\lambda_N + \lambda_b). \tag{10}$$

In Eqs. (9) and (10), an on-off keying modulation format with a duty cycle *q* and a Poisson distribution, which is a common assumption for stochastic links, are assumed. $\lambda_L = \tau\eta G_{LOS}P/E_p$ and $\lambda_N = \tau\eta G_{NLOS}P/E_p$ represent the mean numbers of detected photoelectrons of signal component in each slot for the LOS and NLOS links, respectively. *P* is the output power from Alice and *η* is the receiver efficiency, which are identical for Bob and Eve and are always set at *P* = 1 W and *η* = 0.1 in the calculations. $E_p$ is the energy of one THz photon and *τ* is integration time of the receivers of Bob and Eve. $\lambda_b$ represent the mean number of detected photoelectrons of background radiation component in each slot. In experimental measurement, the THz radiation is converted to direct current (DC) electrical power using a rectifying diode and an antenna, which is based on the conversion of photoelectron to current. So here, we take the photoelectron in consideration in our calculation and the signal to noise ratio (SNR) of receiver for LOS channel can be obtained by dividing $\lambda_L$ by $\lambda_b$.

### 3.1 Secrecy Performance in Rain

In the calculation, outdoor links with carriers at 140, 220 and 340 GHz are considered (which has been achieved in [35-38]) and assumed to be collimated with the same beam width. The FOV angles of Eve for the links are set to be 15° and the receiving areas for Bob and Eve are all 1cm², which are all chosen in the following calculations. To calculate the atmosphere (gaseous) attenuation, we set temperature *T* = 25°C, pressure *P* = 1013 hPa, and relative humidity RH = 97%. When Eve is set at a position of (100m, 10m), the link gain is calculated and shown in Fig. 4. Solid curves represent the evolution of link gain $G_{LOS}$ for the LOS link as a function of rainfall rate. It decreases as the increasing of rainfall rate due to the increasing of attenuation by rain. With the increasing of carrier frequencies, the LOS link gain decreases due to the higher atmospheric attenuation as indicated in Fig. 1. The dashed lines represent the variation of NLOS link gain $G_{NLOS}$. When at 140 GHz, it is smaller than the $G_{LOS}$ at small rainfall rate corresponding to small scattering effect. However, when the rainfall rate increases to 27 mm/hr, both link gain $G_{LOS}$ and $G_{NLOS}$ intersect ($G_{LOS} = G_{NLOS}$). And the $G_{NLOS}$ becomes larger than the $G_{LOS}$ at higher rainfall rate $Rr > 27$ mm/hr due to the larger signal attenuation suffered by the LOS link. For the carriers at 220 GHz, the links gains are smaller than that at 140 GHz and the intersection can be achieved at 4 mm/hr rainfall rate, which indicates a reduced link security because of the larger atmospheric attenuation and scattering effect. For the carrier at a higher frequency (340 GHz), the link gain for LOS and NLOS links at 340 GHz is much smaller than -200 dB and cannot be shown in the plot. So here, we can say that the link at lower carrier frequencies is less vulnerable to eavesdropping attacks in rain when an eavesdropper is positioned at (100m, 10m).

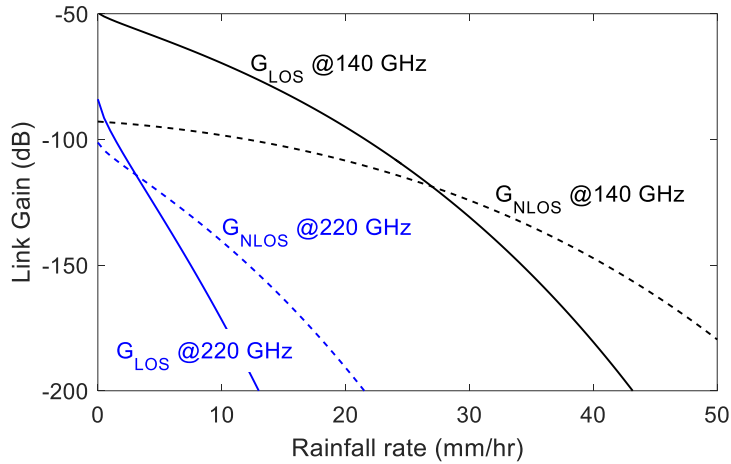

**Figure 4** LOS (solid) and NLOS (dashed) link gain based on Eqs. (2) and (3) with respect to rainfall rate with carriers at 140GHz (black), 220GHz (blue) with Eve at a position of (100m, 10m) and Bob located at (1km, 0m). ($T = 25°C$, $P = 1013$ hPa and RH = 97%)

The secrecy capacity distributions with respect to arbitrary positions of Eve are calculated and shown in Fig. 5 when the rainfall rate $Rr = 10$ mm/hr. The color bar on the right denotes the safe transmission rate $C_s$ in Gbps, which increases gradually as the color changes from blue to yellow. In Fig. 5(a) for the 140 GHz link, the dark blue region represents $C_s = 0$, which means the secure transmission cannot be guaranteed if Eve is located in this region. We would call it as insecure region. For the 220 GHz link in Fig. 5(b), the area of insecure region becomes smaller due to the much higher signal attenuation suffered by the LOS link in rain, which reduces the signal-to-noise ratio (SNR) and further lower the possibility of eavesdropping attacks. This has been demonstrated in [7] for satellite networks. The degradation of SNR could also lead to decreasing of the maximum data capacity (maximum safe data transmission rate) to 90 Gbps, which is lower than that at 140 GHz (around 145 Gbps) in Fig. 5(a). When we set the carrier frequency to 340 GHz as in Fig. 5(c), the insecure region is further reduced and the maximum safe data transmission rate becomes 60 Gbps.

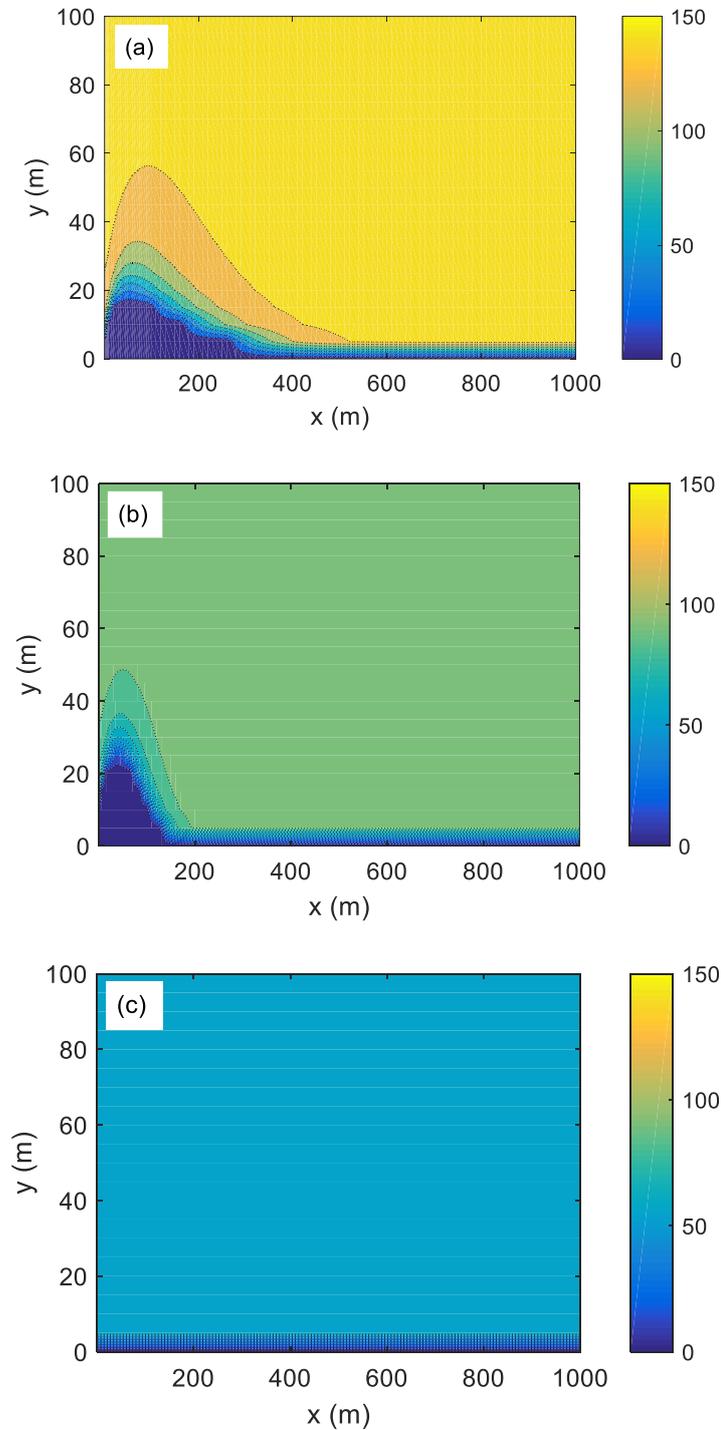

**Figure 5** Secrecy capacity distribution based on Eq. (8) for 2-D positions of Eve when (a) 140 GHz link, (b) 220 GHz link, and (c) 340 GHz link with rainfall rate $Rr = 10$ mm/hr and Bob located at (1km, 0m). ($T = 25°C$, $P = 1013$ hPa and RH = 97%. The color bar denotes the safe transmission rate

in Gbps).

## 3.2 Secrecy Performance in Dry Snow

With the same parameter settings as in Fig. 4, the link gain for LOS and NLOS links in dry snow is calculated and shown in Fig. 6 with Eve set at a position of (100m, 10m), temperature $T = 0°C$, pressure $P = 1013$ hPa and relative humidity RH = 97%. We choose G-M model as snowdrop size distribution. Solid and dashed lines represent evolution of link gain $G_{LOS}$ and $G_{NLOS}$, respectively, which are higher than that in rain due to the smaller signal loss in dry snow. Similar to the trend in Fig. 4, the LOS link gain $G_{LOS}$ decreases as the increasing of snowfall rate (equivalent rainfall rate) due to the increasing of attenuation by snow, but there is no intersection between $G_{LOS}$ and $G_{NLOS}$ when carrier is at 140 GHz, which means the link is always secure over the snowfall rate range of 0-50 mm/hr. However, the intersection (where, $G_{LOS} = G_{NLOS}$) appears when $Rr = 50$ mm/hr for the 220 GHz link and $Rr = 10$ mm/hr for the 340 GHz link due to the more serious link degradation suffered by higher carrier frequencies. This is consistent with the results in rain.

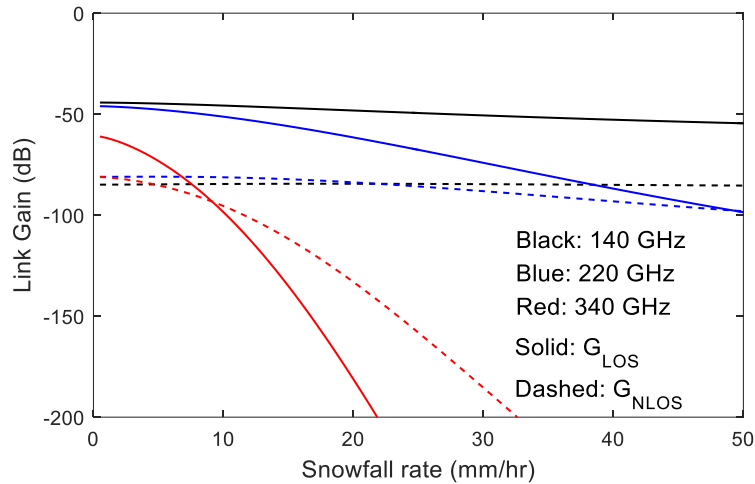

**Figure 6** LOS (solid) and NLOS (dashed) link gain based on Eqs. (2) and (3) with respect to snowfall rate (equivalent rainfall rate) with carriers at 140GHz (black), 220GHz (blue), 340GHz (red) with Eve at a position of (100m, 10m) and Bob located at (1km, 0m). ($T = -1°C$, $P = 1013$ hPa and RH = 97%)

Fig. 7 shows the secrecy capacity distributions with respect to arbitrary positions of Eve in dry snow when the snowfall rate (equivalent rainfall rate) $Rr = 10$ mm/hr. In Fig. 7(a) for the 140 GHz link, the dark blue (insecure) region is much larger than that in Fig. 5(a), even though the maximum safe data transmission rate is still at 145 Gbps. We attribute this to the larger snow particle density under the same fall rate. The snowfall rate is usually represented by equivalent rainfall rate as in [6], so, under the same fall rate of 10 mm/hr, there should be much more dry snow particles than rain particles dropping down, which would lead to more serious scattering effect in dry snow. This can explain why a worse secrecy in dry snow is observed, even though the link signal loss is less as the comparison between Figs. 1 and 2. When the carrier frequency is set at 220 GHz in Fig. 7(b), the insecure region increases and the maximum safe data transmission rate is reduced to 90 Gbps. This is due to the more scattered power from the 220 GHz link than the 140 GHz link which is confirmed in Fig 2(a) and demonstrated in [39]. However, when we increase the carrier frequency to 340 GHz, the area of insecure region is reduced because of the increasing of signal absorption on the LOS link, which leads to a much lower SNR at Bob.

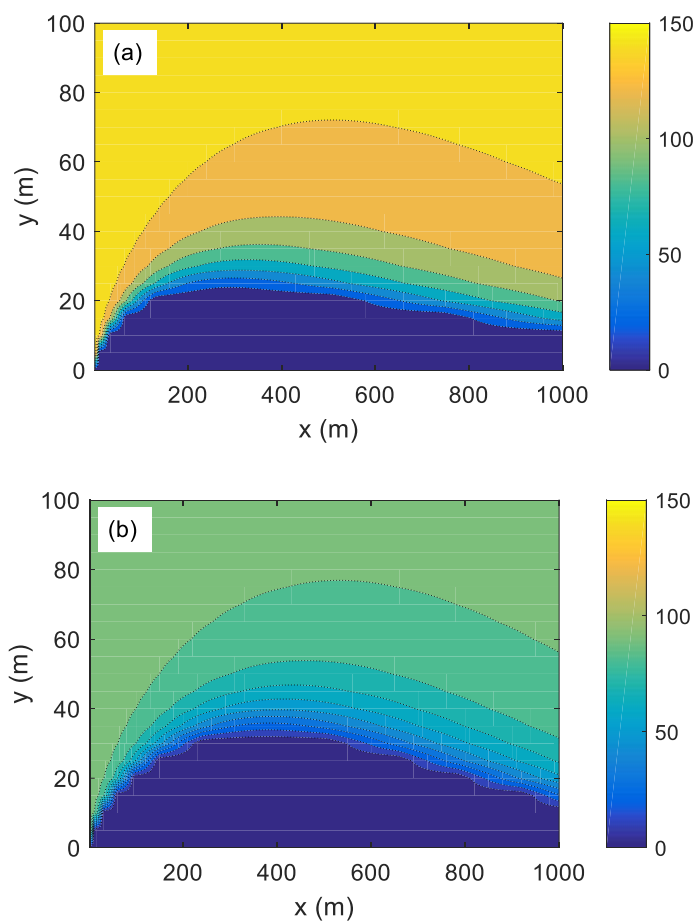

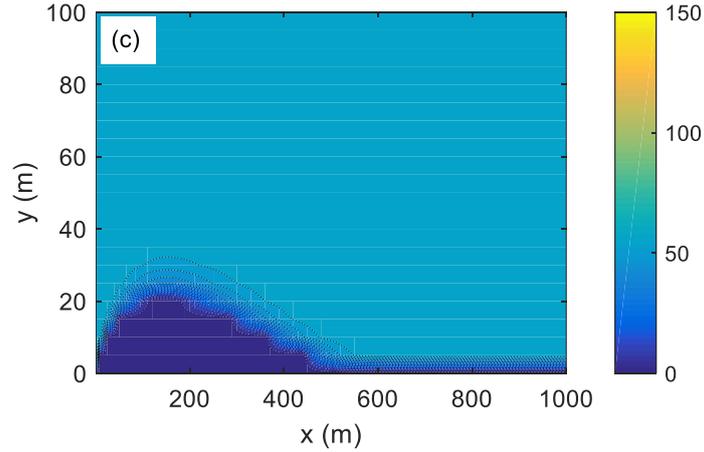

**Figure 7** Secrecy capacity distribution based on Eq. (8) for 2-D positions of Eve when (a) 140 GHz link, (b) 220 GHz link and (c) 340 GHz link with .

snowfall rate (equivalent rainfall rate) $Rr$ = 10 mm/hr and Bob located at (1km, 0m). ($T$ = 25°C, $P$ =1013 hPa and RH = 97%. The color bar denotes the safe transmission rates in Gbps).

### 3.3 Secrecy Performance in Wet Snow

In wet snow with temperature $T = 0$°C, pressure $P = 1013$ hPa and relative humidity RH = 97%, the link gains for Bob and Eve are calculated and shown in Fig. 8 when the G-M raindrop size distribution model is considered again. For carriers at 140 GHz, the LOS link gain $G_{LOS}$ is always higher than the NLOS link gain $G_{NLOS}$ with the snowfall rate up to 50 mm/hr. This means the link at 140 GHz is secure when an eavesdropper set at (100m, 10m). However, this secrecy is decreased when the carrier frequency is changed to 220 GHz, where an intersection ($G_{LOS} = G_{NLOS}$) appears with snowfall rate at $Rr = 41$ mm/hr, which indicates that the link security attack comes when snowfall rate (equivalent rainfall rate) $Rr > 41$ mm/hr. The intersection point for the 340 GHz link comes at an even smaller snowfall rate $Rr = 16$ mm/hr and the link secrecy is much worse.

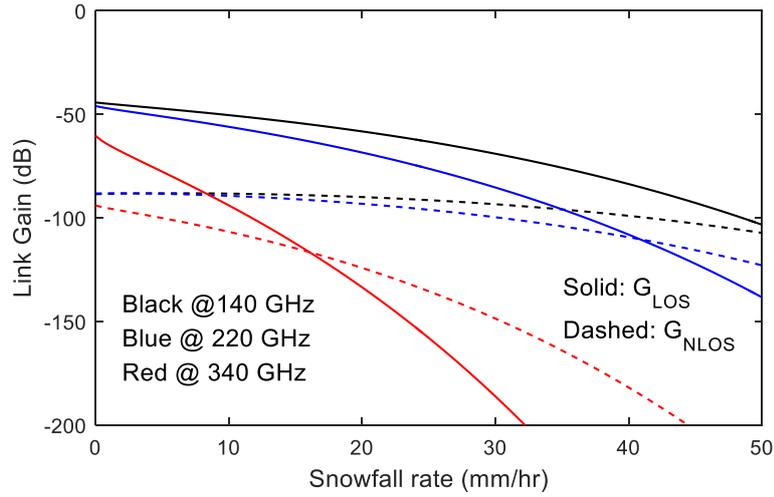

**Figure 8** LOS (solid) and NLOS (dashed) link gain based on Eqs. (2) and (3) with respect to snowfall rate (equivalent rainfall rate) with carriers at 140GHz (black), 220GHz (blue) and 340GHz (red) with Eve positioned at (100m, 10m) and Bob located at (1km, 0m). ($T = 0°C$, $P = 1013$ hPa and RH = 97%)

When we set the snowfall rate $Rr = 10$ mm/hr, the secrecy capacity distributions with respect to arbitrary positions of Eve in wet snow are calculated and shown in Fig. 9. For the 140 GHz link, the insecure region is smaller than that in dry snow in Fig. 7(a) under the same snowfall rate due to less scattering it suffered in wet snow. We also attribute this to the more snow particles in dry snow than in wet snow (where more water contained inside) when under the same equivalent rainfall rate. This phenomenon is also observed for the 220 GHz and 340 GHz links as in Fig. 9(b) and (c), which indicates that the link secrecy performance in wet snow is better than that in dry snow. Besides, with the increasing of carrier frequency from 140 GHz to 220 and 340 GHz, the insecure region decreases and the maximum safe data transmission rate decreases from 145 Gbps to 90 Gbps and 60 Gbps, respectively.

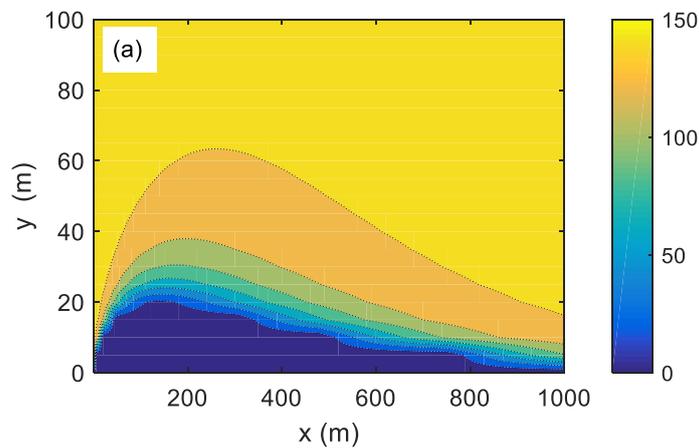

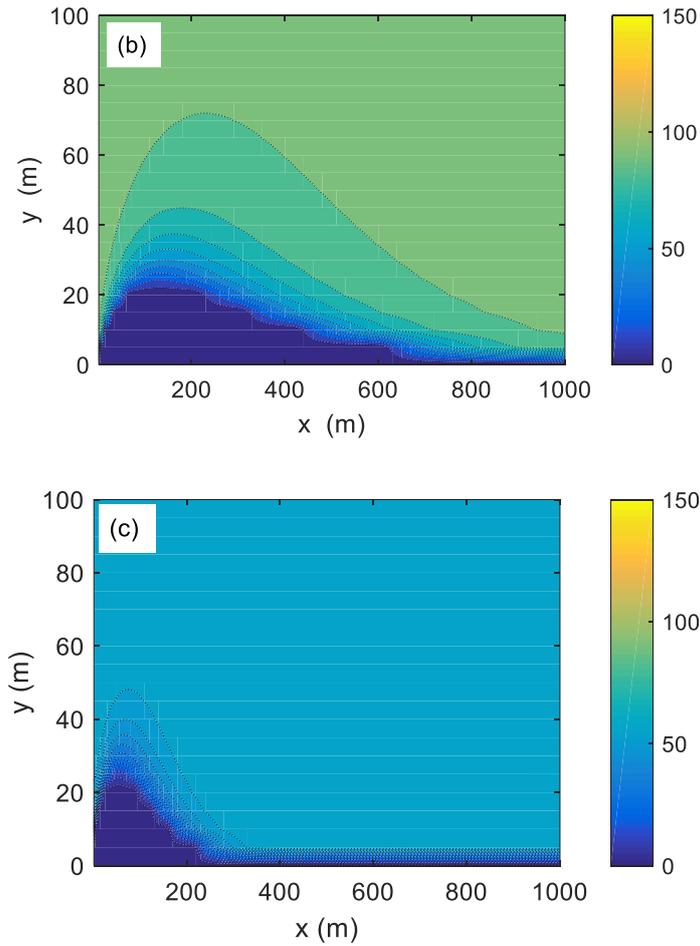

**Figure 9** Secrecy capacity distribution based on Eq. (8) for 2-D positions of Eve when (a) 140 GHz link, (b) 220 GHz link and (c) 340 GHz link with snowfall rate (equivalent rainfall rate) $Rr = 10$ mm/hr and Bob located at (1km, 0m). ($T = 25°C$, $P = 1013$ hPa and RH = 97%. The color bar denotes the safe transmission rates in Gbps).

## 4. Discussion and Conclusions

To further investigate the influence of rain/snow fall rate and receiver sensitivity of Eve, we plot the secrecy capacity of a 220 GHz link. The x-position of Eve is still set at 100m as in Table 1, while its y-position varies from 0m to 50m. In Fig. 10, the secrecy capacity of the link increases significantly when Eve moves away Alice (corresponding to the increasing of y-position). Finally, it would reach to a constant where the maximum secrecy capacity (maximum safe data transmission rate) could be achieved. The vertical part implies a secrecy capacity of 0 Gbps, which means security risks come when Eve is located at this y-position or smaller than it. So we could obtain the insecure region again. In Fig. 10(a), the area of the insecure region decreases for larger rain intensity (rainfall rate), which is opposite to the trend observed in dry (Fig. 10(b)) and wet (Fig. 10(c)) snow. We attribute this to the much weaker scattering effect in rain when compared with absorption as in Figs. 1 and 2. With the increasing of rain intensity, the higher signal loss suffered by the LOS link in rain and the smaller scattered power for the NLOS link would reduce the possibility of eavesdropping attacks more and more. While in dry and wet snow, the scattering effect is close or even more serious than the absorption, which indicates that the NLOS link gain would decrease much more slowly than the LOS link gain as what we have observed in Figs. 4, 6 and 8.

According to Eqs. (8), (9) and (10), it is straightforward that the link secrecy performance could be increased for larger receiver sensitivity (smaller SNR) of Bob and smaller receiver sensitivity (larger SNR) of Eve. To further confirm this, we predict the secrecy capacity evolution with SNR of Eve changes from 0 dB to 6 dB. With the decreasing of Bob' receiver sensitivity, the insecure region is reduced and the maximum secrecy capacity is increased significantly.

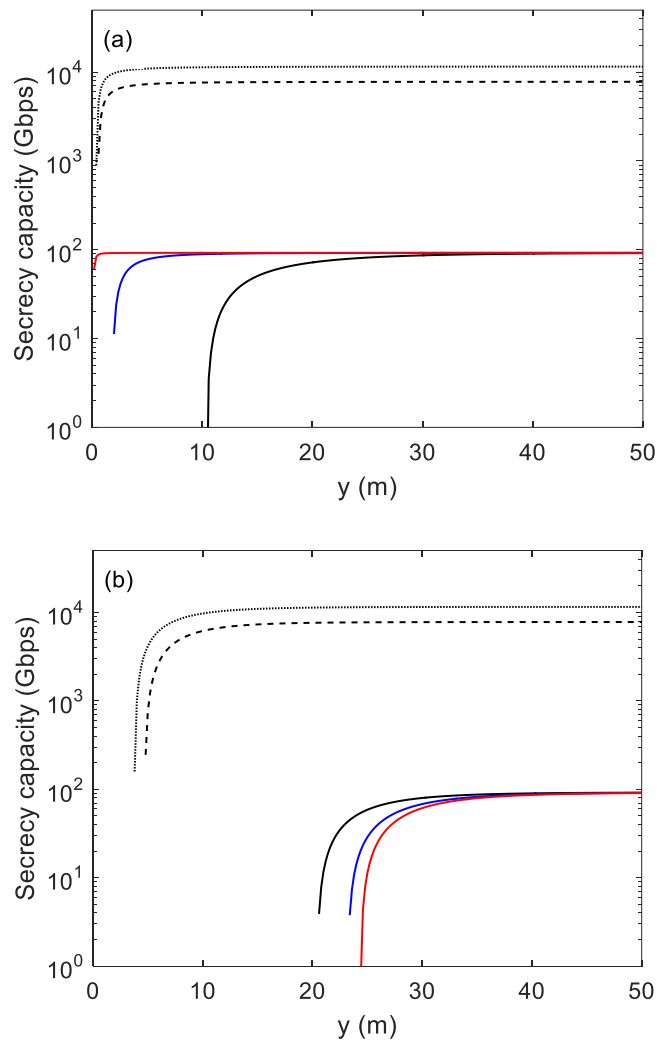

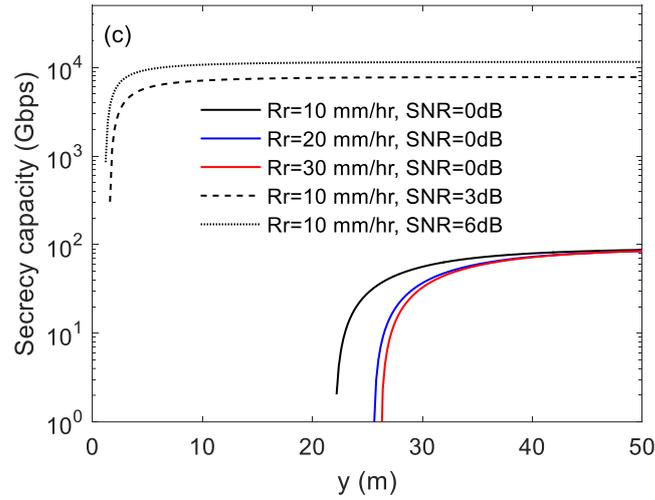

**Figure 10** Secrecy capacity based on Eq. (8) with respect to y-position of Eve in falling (a) rain, (b) dry snow and (c) wet snow with a carrier frequency at 220 GHz and an x-position of 100m. *Rr* is the equivalent rainfall rate in mm/hr, the SNR is signal-to-noise ratio of Eve.

As a whole, in this work, we investigate the secrecy performance of a LOS THz link in rain and snow when an eavesdropper locates outside of the legitimate link path and tries to collect and decode the scattered signal. A theoretical model combining Mie scattering theory, stochastic models and gaseous attenuation is proposed to estimate the link gain (of legitimate and eavesdropping links) and secrecy performance. The calculation results show that the absorption loss in rain is much larger than the scattering loss, while the difference between both is much smaller in wet snow and the scattering loss is even larger in dry snow. The link secrecy capacity distribution in rain and snow is calculated and insecure region is obtained. The maximum safe date transmission rate of a THz link could be improved by decreasing carrier frequency and/or receiver sensitivity of Eve, while rain/snow intensity has no influence on it. Insecure region is inversely proportional to receiver sensitivity of Bob in falling rain/snow. Smaller insecure region is observed in rain than in snow under identical equivalent rainfall rate, and its area is negatively related to the rain intensity due to the much higher absorption effect than scattering. In other words, higher physical player security could be achieved in heavier rain, while in dry and wet snow, larger snow intensity would reduce the link security.

**Acknowledgements**

This research was supported by National Natural Science Foundation of China (No. 6207106) and Beijing Institute of Technology Research Fund Program for Young Scholars (No. 2020CX04087).

## Disclosures

No conflicts of interest.